\documentstyle[12pt]{article}
\title{Effect of the hole doping into the Haldane-gap state
on the one-dimensional $S=1$ $t$-$J$ model}
\author{Yoshihiro Nishiyama and Masuo Suzuki \\
{\it Department of Physics, University of Tokyo}\\
{\it Hongo 7-3-1, Bunkyo-ku, Tokyo 113, Japan}}
\date{(Received \hspace*{50mm})}

\begin{document}
\begin{normalsize}

\maketitle

\begin{abstract}
The ground state and excitation spectra of
the $S=1$ Heisenberg spin chain
with hole hopping
are investigated
by means of the numerical-diagonalization method and
by the help of the Zhang-Arovas
picture.
As for the charge sector, two phases are found.
It is shown that
in the weak-hopping region, a phase separation occurs,
while in the strong-hopping region, the system is in
the Tomonaga-Luttinger liquid phase.
Irrespective of the presence of the spin gap, the present phase diagram
is similar to that reported for the $S=1/2$ $t$-$J$ chain with gapless
spin excitation rather than to that for
the magnetically-frustrated $t$-$J$ chain with
spin gap.
For the spin sector, the string correlation is found to develop
withstanding the hole doping except in a region, whose
region is characterized
by a generalized string correlation.
This remains an open question to be studied in future.
Submitted to Physica B
\end{abstract}

\section{Introduction}
Since the existence of a spin gap on high-$T_{\rm c}$ Cu-oxides was reported
\cite{Rossat91a,Rossat91b},
electron systems with a spin gap have been studied in the context
of the superconductivity.
In order to cause a spin gap, a number of
generalizations of interactions have been made by
restricting the dimensionality within
one dimension:
the $t$-$J$ model with alternating magnetic interaction
\cite{Imada93},
the $t$-$J$ model with next-nearest-neighbor magnetic frustration
\cite{Ogata91b}, and the double-chain $t$-$J$ model \cite{Dagotto92}.
In particular, the double-chain model
has lately attracted considerable attention because of the discovery
of the corresponding materials \cite{Johnston87,Takano92};
see the articles
\cite{Scalapino95,Hiroi95,Dagotto95}
for a review.

Amongst such spin-gap states, the Haldane-gap state has been studied
extensively.
The spin gap
was conjectured to open for {\it integer}-spin antiferromagnetic
Heisenberg
chains \cite{Haldane83a,Haldane83b}.
The conjecture was partly confirmed both theoretically \cite{Kolb83}
and experimentally \cite{Buyers86}.
Recently, carrier doping to the Haldane-gap material was tried
\cite{Batlogg94,Ramirez94}.
The doping causes the reduction of resistivity and creates
states within the Haldane gap
\cite{Di94}.
There are several theoretical proposals to describe such
experimental results
\cite{Penc95,Dagotto95b,Koshibae95}.
These proposals are based upon some model
Hamiltonians in which each hole is regarded to have the inner freedom
of the spin $S=1/2$, and propagates along an
$S=1$ spin chain.
The hopping amplitude of carriers is, however, estimated
to be so small that
only the considerations upon the
impurity problems could explicate essential properties
\cite{Kaburagi94,Sorensen95,Lu95,Tonegawa95}.

In the present paper, we investigate the following one-dimensional
$S=1$ $t$-$J$ model under the periodic-boundary condition:
\begin{equation}
{\cal H}=
-t \sum_{i=1}^{L} \sum_{\sigma=-1,0,1}
(b_{i\sigma}^\dagger b_{i+1,\sigma} + {\rm h.c.} )
+J\sum_{i=1}^{L} {\bf S}_i \cdot {\bf S}_{i+1}.
\label{Hamiltonian}
\end{equation}
The operators $b_{i\sigma}$ and $b_{i\sigma}^\dagger$
obey the hard-core-boson statics
\begin{eqnarray}
b_{i\sigma}b_{i\sigma'}&=&0, \nonumber \\
\{ b_{i\sigma},b^\dagger_{i\sigma'} \}&=&\delta_{\sigma\sigma'}, \nonumber \\
\left[ b_{i\sigma},b_{i\sigma'} \right] &=& 0 \ (i\ne j), \nonumber \\
\left[ b_{i\sigma},b^\dagger_{i\sigma'} \right] &=& 0 \ (i\ne j).
\end{eqnarray}
The operators $\{ {\bf S}_i \}$ are  expressed as
\begin{equation}
{\bf S}_i=
\left(b^\dagger_{i,1}\ b^\dagger_{i,0}\ b^\dagger_{i,-1}\right)
{\bf s}
\left(
\begin{array}{c}
b_{i,1} \\
b_{i,0} \\
b_{i,-1}
\end{array}
\right),
\end{equation}
where the matrices ${\bf s}=(s^x,s^y,s^z)$ represent the $S=1$ spin
operators.
This model may be too simplified to describe the above magnetic material.
Apart from the practical interest, however, the present Hamiltonian
(\ref{Hamiltonian}) may be useful for studying essential effects of a spin
gap to phase diagrams.
This is quite accordant with our motivation explained at the beginning of the
present paper.

For some limiting cases, some properties of the ground state of the
Hamiltonian (\ref{Hamiltonian}) are known as follows.
At the filling $n=1$, the model reduces to the $S=1$ spin system;
the system has the so-called Haldane gap,
$\Delta E\approx0.4105 J$ \cite{White93,Golinelli94}.
The two-particle problem corresponding to the low-density limit,
namely $n=\frac2L$,
can be solved exactly;
two particles make a bound state
in the region $J/t>1$,
while
they are deconfined
in the region $J/t<1$.
Hence, it is expected that in the region
$\displaystyle{J/t\mathop{>}_{\sim}1}$
particles
are apt to form an island of particles for
arbitrary filling $n$; the so-called phase separation takes place.
Hence,
the characteristics of the system is rather simple in this phase-separated
region.
Notice that this situation is relevant to the
above material.
We stress here that the present paper mainly concerns the liquid phase
$J/t\displaystyle{\mathop{<}_{\sim}}1$, in which
the properties of the fluid are nontrivial.

The present paper is organized as follows.
In the next section, we explain the Zhang-Arovas argument \cite{Zhang89}
which is also applicable
to the present problem (\ref{Hamiltonian}).
This yields a rough feature of the phase diagram in our system, and predicts
the existence of a spin gap withstanding the hole doping, and so forth.
Referring to these predictions, we present in \S 3 numerical results for the
model (\ref{Hamiltonian})
both on the charge and spin sectors.
It is shown that such a new phase as is not predicted
from the Zhang-Arovas
argument spreads
in a low-density region.
In addition to this, we report that the present system in the metallic phase
is not essentially affected by the presence of the Haldane gap.
The phase diagram resembles that of the $S=1/2$ $t$-$J$ model
\cite{Ogata91a} without
spin gap rather than that of the magnetically-frustrated
$t$-$J$-$J'$ model
\cite{Ogata91b} with a spin gap.
The present spin-liquid
state seems to be rather unique compared to other
spin-liquid states.
In the last section, we give a summary of the present paper.

\section{Zhang-Arovas picture for the present model}
In this section, we explain the Zhang and Arovas picture
relevant to the present system
\cite{Zhang89}.
They discussed the following system;
\begin{equation}
{\cal H}=-t\sum_{i=1}^{L} \sum_{\sigma=-1,0,1}
(b^\dagger_{i,\sigma} b_{i+1,\sigma} + {\rm h.c.})
         +J\sum_{i=1}^{L} \left({\bf S}_i \cdot {\bf S}_{i+1}
         +\frac13 ({\bf S}_i \cdot {\bf S}_{i+1})^2 \right).
\label{ZA}
\end{equation}
The magnetic interaction including
the biquadratic term is precisely the same
as that in the AKLT model
\cite{AKLT87,AKLT88}
, which can be solved exactly at the ground state

First, we consider the impurity problem $t=0$.
At the condition $t=0$,
the model reduces to a lot of
AKLT sectors under the {\it open}-boundary condition;
the system is cut off at the positions of holes to form independent
finite AKLT
spin
chains.
On the other hand,
the ground state of the AKLT model is called the valence-bond solid
(VBS) state.
The VBS state for the system of length $l$ under the
open-boundary condition
is given as \cite{Arovas88}
\begin{equation}
|{\rm VBS}\rangle= (a^\dagger_1)^p(b^\dagger_1)^{1-p}
\left( \prod_{j=1}^{l-1}
    (a^\dagger_j b^\dagger_{j+1}-b^\dagger_j a^\dagger_{j+1})
\right)
                   (a^\dagger_l)^q(b^\dagger_l)^{1-q}
                   |0\rangle,\ (0\le p,q \le 1),
\label{VBS}
\end{equation}
where the $S=1$
spin operators are expressed in terms of the Schwinger representation
\begin{equation}
S^+_i=a^\dagger_i b_i,\ S^-_i=a_i b^\dagger_i,
\ S^z_i=\frac{a^\dagger_i a_i-b^\dagger_i b_i}{2}\
{\rm and}\ a^\dagger_i a_i+b^\dagger_i b_i=2.
\end{equation}
The formula (\ref{VBS}) can be interpreted such that a valence bond
is formed over each bond in a systematic manner.
The edge freedom of $p$ and $q$ in eq. (\ref{VBS})
can be expressed in terms of the following different representation.
The edges holding hole(s) between them
can be classified in terms of the total magnetization of these edges.
The magnetization ${\bf S}^2=0$ implies that a singlet valence bond is
formed over the hole(s),
while the magnetization ${\bf S}^2=1(1+1)$ implies that a triplet
bond is formed;
see Fig. 1.
Once the hopping amplitude $t$ is assumed to be nonvanishing,
this expression becomes
essential.
Hence, it is consequently found that
there are two types of holes that can be classified with respect
to the inherent
bond; we call the former (latter) `singlet hole' (`triplet hole'),
hereafter.

Second, we consider the system (\ref{ZA}) with a finite
hopping term $t$.
It was reported that the singlet-hole state with the wave number $k$
is an exact eigenstate, while the triplet-hole state is not \cite{Zhang89}.
Hence,
the Hamiltonian for the singlet holes can be given exactly
in terms of the spinless-fermion operators;
\begin{equation}
{\cal H}=-t\sum_i \left( c^\dagger_{i}c_{i+1}+{\rm h.c.} \right)
         -\frac{2}{3}J\sum_i c^\dagger_i c_i c^\dagger_{i+1}c_{i+1}
\label{fermion}
\end{equation}
Hence, the band of one singlet hole is given by
$\epsilon_k=-2t\cos k$.
That for one triplet hole is approximately estimated as
$\epsilon_k=\frac{2t}{3}\cos k$
\cite{Zhang89}.
Hence, the singlet-hole state is concluded to be the ground state
in the presence of the hopping term.
The Hamiltonian (\ref{fermion}) can be transformed
in terms of the Jordan-Wigner transformation
into the following
ferromagnetic Heisenberg chain:
\begin{equation}
{\cal H}=-4t\sum_i \left( \sigma^x_i \sigma^x_{i+1}+
                          \sigma^y_i \sigma^y_{i+1}+
              \frac{J}{3t}\sigma^z_i \sigma^z_{i+1} \right),
\label{ferro}
\end{equation}
where the Hilbert space is restricted within
$\sum_i \sigma^z_i=2N-L$; $N$ denotes the number of the particles.
Now, we are in a position to grasp a rough feature of the phase diagram
of the present system with singlet holes.
In the $XY$ limit $J/t\to0$, the ground state is in the
$XY$ phase, {\it i.e.}, the Tomonaga-Luttinger liquid phase.
In the Ising
limit $J/t\to\infty$, the state is phase-separated: the
positive-magnetization sector and the negative-magnetization sector
extend separately.
Along the conditions $\sum_i \sigma^z_i=\pm L \mp 2$, {\it i.e.,} the high- or
low-density limits,
the state  shows the confinement-deconfinement transition
at $J/t=3$.
Note that this transition point coincides the isotropic condition of
the spin system (\ref{ferro}).
Keeping in mind of the particle-hole symmetry
$c_i \leftrightarrow (-1)^i c^\dagger_i$ of the Hamiltonian
(\ref{fermion}),
the phase diagram of the singlet-hole state
may be drawn schematically as in Fig. 2.

Lastly, we explain the properties of the spin sector.
If the singlet-hole state is realized at the ground state,
the string correlation
\cite{den89,Tasaki91}
\begin{equation}
{\cal O}^z_{\rm string}(j-i)=
\langle S^z_i {\rm e}^{{\rm i}\pi\sum_{k=i}^{j-1} S^z_k}
        S^z_j \rangle
\label{string}
\end{equation}
develops withstanding the hole doping.
The reason is as follows.
We assume that the singlet-hole state should be an ground state.
Then, focusing  only on the $S=1$ particles, {\it i.e.,} disregarding
the holes, the spin sector is expressed in terms of the
VBS state;
a valence bond is formed over each neighboring $S=1$ particles.
On the other hand, the string correlation (\ref{string})
is known to develop on the VBS state.

The existence of the developing string correlation
can be interpreted as the existence of a spin gap
because of the following reason.
This existence implies that the state should be relevant to the VBS state.
A finite energy is required to excite the state, whose energy corresponds to
exiting
one of the valence bonds to a triplet bond.
Hence, in the present paper, we study the string correlation rather than
 the spin gap directly.

\section{Numerical Results and Discussions}
In this section, we investigate the ground-state properties of the system
(\ref{Hamiltonian}) by means of the exact-diagonalization method.
The results are discussed by the help of the Zhang-Arovas picture
explained in the previous section.
Phase diagrams are given both for the charge and spin sectors.

\subsection{Elementary excitations}
In this subsection, we show dispersion relations both for the phase-separated
region
$J/t\displaystyle{\mathop{>}_{\sim}}1$
and liquid region $J/t\displaystyle{\mathop{<}_{\sim}}1$,
before estimating the precise phase boundary.
Dynamical structure factors are evaluated so that
they can yield the respective elementary excitations of the charge and spin
sectors. This separation enables us to analyze the charge sector
in terms of the Tomonaga-Luttinger theory;
see the article \cite{Schulz91} for a review.

First, in Fig. 3 (a), we present the dispersion relation for the system with
$L=14$, $J/t=10\gg1$ and $n=13/14$ which is a
single-hole system in the phase-separated
region.
The low-lying bands originate in
the single-hole
dispersions.
Because of the degeneracy of the band with respect to the total
magnetization $\sum_i S^z_i$, see Fig. 3 (a), the wider (narrower)
band is seen to correspond
to singlet (triplet) one.
According to the Zhang-Arovas picture explained in the previous section,
the band width of the singlet hole is
approximately three-times wider than
that of the triplet one.
The dispersions in Fig. 3 (a)
actually agree with this prediction.
As for the spin sector,
we find that the spin gap opens independently of the wave number $k$,
$\Delta E \sim J$.
The reason of this independency is due to the condition $J/t\ll1$, where
a hole can hardly propagate and can be regarded as an impurity.
As a consequence, the state is independent of the wave number.

Second, in Fig. 3 (b), we show the dispersion relation for the system with
$L=14$, $J/t=1$ and $n=12/14$ which is a system with a
hopping amplitude
comparable to the magnetic interaction.
The structure of elementary excitations is no more manifest.
The spin and charge elementary excitations are separated with use of
the following analysis.

We employed the dynamical structure
factor $S(k,\omega)$ both for the spin and charge sectors,
\begin{equation}
\label{charge}
S_{\rm charge}(k,\omega)=\sum_j
 \langle 0 | n_{-k} |j\rangle \langle j| n_k| 0 \rangle
 \delta (\omega - E_j + E_0),
\end{equation}
and
\begin{equation}
\label{spin}
S_{\rm spin}(k,\omega)=\sum_j
 \langle 0  S^z_{-k} |j\rangle \langle j| S^z_k | 0 \rangle
 \delta (\omega - E_j + E_0),
\end{equation}
where
$n_k=L^{-\frac12}\sum_{i,\sigma} {\rm e}^{-{\rm i}kr_i}b^\dagger_{i,\sigma}
b_{i,\sigma}$ and
$S^z_k=L^{-\frac12}\sum_{i,\sigma} {\rm e}^{-{\rm i}kr_i}S^z_{i}$.
The factors show $\delta$-function peaks at the levels which have the
above particular transition probabilities.
They are evaluated with use of Mori's continued-fraction-expansion formalism
\cite{Mori65,Gagliano87}.
The factors (\ref{charge}) and (\ref{spin})
for the system with $L=14$, $J/t=1$ and $n=12/14$ are shown in
Fig. 4 (a) and (b), respectively,
where the peaks are broadened with the parameter $\eta=0.04$
such as $\delta(x)=-\frac{1}{\pi}{\rm Im} \frac{1}{x+ {\rm i} \eta}$.
As was discussed in the previous section, the charge sector
in the region
$J/t \displaystyle{\mathop{<}_{\sim}}1$
is expected to be described in terms of the Tomonaga-Luttinger theory.
It is crucial to estimate the sound velocity $v_{\rm c}$ that is defined
with respect to the low-lying
dispersion relation $\omega=v_{\rm c} k$ ($k\sim0$).
Through collating Fig. 4 (a) with Fig. 3 (a),
it can be seen that the sound velocity for the charge elementary excitation
should be estimated with respect to the level A; $v=(E_{\rm A}-E_0)/
\left(\frac{2\pi}{L}\right)$.
The estimated sound velocity for the system with $L=14$, $n=12/14$
and $J/t$ varied
is shown in Fig. 5.

As for the spin elementary excitation depicted in Fig. 4 (b),
the first exited state
locates at the wave number $k=k_{\rm F}=n\pi$.
This location  coincides with that for the $S=\frac12$ $t$-$J$ model,
whose model was, however, shown to have a {\it gapless}
spin excitation spectrum by the use of the rigorous solution along the
tractable line $J/t=2$ \cite{Bares90}.
The spin excitation of the present model
is shown to have a spin gap in the subsection 3.3.

\subsection{Phase diagram of the charge sector}
In this subsection, we investigate the phase diagram of the charge sector.
It is analysed in term of
the Tomonaga-Luttinger liquid theory,
where the estimation of
the sound velocity $v_{\rm c}$ given in the previous subsection
plays an essential role.

The exponent $K_\rho$ which governs the Tomonaga-Luttinger liquid state
can be expressed by \cite{Haldane81,Schulz90},
\begin{equation}
K_\rho=\frac\pi2 v_{\rm c}n^2\kappa,
\label{exponent}
\end{equation}
where $\kappa$ denotes the compressibility
$\kappa=\frac{1}{Nn}\frac{\partial^2 E_0}{\partial N^2}$.
This compressibility is estimated by means of the difference method as
\begin{equation}
\kappa=\frac{L}{N^2}
\frac{E_0(N+2)+E_0(N-2)-2E_0(N)}{4}.
\end{equation}
Every correlation exponents of the
charge sector are governed in terms of the single parameter
$K_\rho$; for example, the charge-density wave and the pairing correlations
obey the following power-law
\begin{eqnarray}
\label{CDW}
\langle O^\dagger_{\rm density}(r) O_{\rm density}(0)
\rangle &\propto& 1/r^{K_\rho}, \\
\label{pair}
\langle O^\dagger_{\rm pair}(r) O_{\rm pair}(0)
\rangle &\propto& 1/r^{1/K_{\rho}},
\end{eqnarray}
\begin{center}
$\left(
O_{\rm density}(i)=\sum_\sigma b^\dagger_{i\sigma}b_{i\sigma}\
O_{\rm pair}(i)=\sum_\sigma b^\dagger_{i\sigma}b^\dagger_{i+1,\sigma}
\right),$
\end{center}
respectively.
At $K_\rho=1$, the above correlations (\ref{CDW}) and (\ref{pair})
decay with the same exponent.
The former (latter) should be dominant, if the liquid is ``repulsive
(attractive).''
Hence, it is seen that
the magnitude of the exponent $K_\rho$ indicates a
degree of  the attraction among the particles and
that the particles are ``noninteractive'' at $K_\rho=1$.
This interpretation will be found to be valid
through referring to the exact estimation of $K_\rho$ for various models
\cite{Frahm90,Kawakami90}.

In Fig. 6, we presented the estimation of $1/K_{\rho}$ for the system with
$L=14$, $n=12/14$ and $J/t$ varied.
In the weak-interaction limit $J/t\to0$, only concerning the charge sector,
the present system converges to the
spinless-hard-core gas.
For such a system, the exponent $K_\rho$ is given by $K_\rho=0.5$ exactly.
This is quite consistent with the present numerical estimation in Fig. 6.
As the magnetic interaction $J/t$ is increased, the exponent
$K_\rho$ increases.
In the region $1\displaystyle{\mathop{<}_{\sim}}%%%
J/t\displaystyle{\mathop{<}_{\sim}}1.6$,
the exponent is estimated as $K_\rho>1$.
Hence, the liquid is attractive; the paring correlation
(\ref{pair}) is dominant in this region.
On the other hand,
the exponent $K_\rho$ is negative
in the region $\displaystyle{J/t\mathop{>}_{\sim}1.6}$.
Therefore, the compressibility $\kappa$
is also negative in this region; see eq. (\ref{exponent}).
Hence, the gas is unstable; we can regard the system as being phase-separated
in this region.
Note that these features are consistent with the argument explained
in \S 2.

With use of
the similar data for various values of the
filling $n$, we obtained the phase diagram for
the charge sector, see Fig. 7.
As for the charge sector,
two qualitatively-different phase diagrams
have been reported so far (see Fig. 8
(a) and (b)).
The phase diagram of the present model belongs to the type depicted
in Fig. 8 (a).
Notice that the phase diagram depicted in
Fig. 2 which is given according to the Zhang-Arovas argument
also belongs to the type of Fig. 8 (a).
The phase diagrams (a) and (b) were reported to hold for the one-dimensional
$S=1/2$ $t$-$J$ model \cite{Ogata91a} and for the $t$-$J$
model with the next-nearest-neighbor
magnetic interaction $J'=0.5J$ \cite{Ogata91b}, respectively.
The former model shows a gapless spin excitation.
On the other hand,
the latter model is expected to have a spin gap near the filling $n\sim1$,
because the model converges to the Majamder-Ghosh model
\cite{Majumdar69} in the limit  $n\to1$.
In the parameter region with $n\sim1$ and
$\displaystyle{J/t\mathop{<}_{\sim}1}$, the diagram Fig. 8
(a) shows
that the liquid has the exponent $K_\rho=0.5$,
while the diagram (b) shows that the liquid is either phase-separated or
has the exponent $K_\rho\gg1$.
Hence, the liquid of the
latter case is strongly attractive one. The reason has attributed
to the presence of a spin gap, so far.

The present model, however, belongs to the type (a) in spite of the
presence of a spin gap, called the Haldane gap.
It suggests that the RVB {\it pattern} is crucial
for enhancing the pairing correlation;
the present RVB pattern belongs to the type
called the VBS state (see the previous
section).

\subsection{Spin sector}
In this  subsection, we concentrate on the spin sector of the present model
(\ref{Hamiltonian}).
The development of the string correlation (\ref{string})
indicates the presence of
a short-range RVB state which is essentially the same as the VBS state;
the state is thus concluded to have a spin gap corresponding to exiting
one of the valence bonds to a triplet.
Physical properties of the liquid phase are of greater interest, because
the phase-separated particle system is no more than an island of the
$S=1$ spin chain in a sea of holes.

In Fig. 9, we have
plotted the string correlation ${\cal O}^z_{\rm string}(L/2)$
for the systems with $J/t=1.0$ and $L,\ n$ varied.
As is expected, the correlation decreases with the filling $n$
decreased.
The plot, however, does not show clearly
whether it develops or not.
In order to see it definitely, we calculated the Binder parameter
\cite{Binder81a,Binder81b,Hatano94}
for the string corelation;
\begin{eqnarray}
U(L)&=&1-\frac{\langle O^4 \rangle}{3 \langle O^2 \rangle^2},\\
O&=&\frac 1 L \sum_i S^z_i{\rm e}^{{\rm i}\pi \sum_{k=1}^{i-1} S^z_k}.
\nonumber
\end{eqnarray}
With the system size increased,
the Binder parameter increases if the corresponding correlation
is developing.
We show the plot
for the systems with $J/t=0.2$, $1.2$, $2.2$ and $L,\ n$
varied (see Fig. 10 (a), (b) and (c), respectively).
At $J/t=0.2$, the string correlation develops withstanding the doping.
Holes coherently move in the VBS state without disturbing it.
Hence, at $J/t=0.2$ we can conclude that the singlet-hole picture is valid at
the ground state.
The above behaviour of the spin gap is in contrast with that for the
frustrated $S=1/2$ $t$-$J$ chain \cite{Ogata91b}.
For this model, it was reported that the spin gap closes in an over-doped
region $n\displaystyle{\mathop{<}_{\sim}}\frac34$.
At $J/t=1.2$, in the low-density region $n\sim0$, the correlation is found to
be of short range. This
phase is absent in the Zhang-Arovas picture introduced
in the previous section.
The nature of this new phase is discussed later.
The correlation is seen to develop
at $J/t=2.2$, which is rather strong.
It is quite natural, if we notice that the system
is
phase-separated  (see the phase diagram depicted in Fig. 7).

With use of similar data for various values of $J/t$,
we draw schematically the transition line in Fig. 11.
As is discussed in \S 1, the two-particle system can be solved exactly;
there exists a confinement-deconfinement transition at $J/t=1$.
We have utilized this additional fact for drawing the transition line
in Fig. 11.

In order to see the nature of the above new phase, we generalize the
correlation
(\ref{string}) to the following form
\cite{Oshikawa92},
\begin{equation}
{\cal O}^z_{\rm string}(|i-j|,\theta)=
\langle S^z_i {\rm e}^{{\rm i}\theta\sum_{k=i}^{j-1} S^z} S^z_j \rangle.
\end{equation}
This has been
successfully applied for the spin-$S$ antiferromagnetic Heisenberg
chains \cite{Oshikawa92,Totsuka95,Nishiyama95}
and the $S=1$ spin chain with the alternating interaction
\cite{Oshikawa92,Totsuka95b}.

For the $S=1$ spin chain with the alternating interaction,
the following was reported.
The correlation at $\theta=\pi$ is maximal in the Haldane phase,
while the correlation at $\theta=\pi$ is vanishing
and that  around $\theta=\pi/2$ is maximal in the dimer phase.
In Figs. 12 (a) and (b),
the correlation ${\cal O}^z(L/2,\theta)$ is plotted for the systems
with (a) $L=14$, $J/t=0.5$ and $n=12/14$ and
(b) $L=14$, $J/t=1.2$ and $n=4/14$,
respectively.
In Fig. 12 (a), a maximum locates at $\theta=\pi$, while in Fig. 12 (b),
a maximum locates about at $\theta=\pi/2$.
The former actually confirms that the state is in the Haldane phase.
On the other hand, the latter implies that the state is, somehow,
in the dimer phase;
the $S=1$ particles may make a singlet bound states in pair.
A similar state is reported in the low-density region for the $S=1/2$ $t$-$J$
chain \cite{Ogata91a,Hellberg93}.
In the present model, however, the region spreads over a large region.
It is noteworthy that the state has a spin gap corresponding to exiting one
of the singlet bound pairs to a triplet pair.
As a consequence, all the phases that appear in Fig. 12 are
found to show a spin gap.

\section{Summary}
The ground state of the
one-dimensional $S=1$ $t$-$J$ model (\ref{Hamiltonian}) is investigated.
In the weak-magnetic-interaction region
$\displaystyle{J/t\mathop{<}_{\sim}1.6}$,
the ground state is in the liquid phase,
while
the system is phase%
-separated
in the strong-interaction region
$\displaystyle{J/t\mathop{>}_{\sim}1.6}$.
In spite of the presence of the spin gap, the phase diagram of the charge
sector belongs to the type that was reported for the one-dimensional
$S=1/2$ $t$-$J$ model rather than that for the frustrated $t$-$J$
model with a spin gap.
Although the presence of a spin gap has been speculated as a sign of strong
attraction among the particles,
the present result suggests that the {\it pattern} of the RVB formation is
rather crucial.
As for the spin sector in the liquid region, in addition to the phase
conjectured by Zhang and Arovas, a new phase is found to appear.
The phase spreads in the low-density region, and the string order is
disturbed. The disturbed state, however, possesses the generalized string
correlation of $\theta=\frac\pi2$. It suggests that the $S=1$ particles
make bound states in pair.
A similar phase is observed in the $S=1/2$ $t$-$J$ model, in whose model,
however,
it expands in a very limiting region \cite{Ogata91a,Hellberg93}.
It turns out consequently that a spin gap outlives to
open in the whole region.
The present spin-liquid state seems to be rather unique
compared to other spin-liquid state.

\section*{Acknowledgement}
Our computer programs
are partly based on the subroutine package "TITPACK Ver. 2"
coded by Professor H. Nishimori.
The numerical calculations were performed on the super-computer HITAC
S3800/480 of the computer centre, University of Tokyo, and on the
work-station HP Apollo 9000/735 of the Suzuki group, Department of
Physics University of Tokyo.

\newpage
{\bf Figure captions}

Fig. 1: Schematic drawing of the generalized
VBS state in the presence of holes.
Over the hole(s), either a singlet or a triplet bond is
supposed to be formed.

Fig. 2: Schematic drawing of the phase diagram of the Hamiltonian
(\ref{ZA}) due to the Zhang-Arovas argument.

Fig. 3 (a) (b): Dispersion relations for the systems
with (a) $L=14$, $J/t=10$ and
$n=13/14$, and (b) $L=14$, $J/t=1$ and $n=12/14$, respectively.
The symbols $+$, $\times$ and $\diamond$ denote the levels
of the quantum number $\sum_i S^z_i=0$, $\pm1$ and $\pm2$,
respectively.

Fig. 4 (a) (b): Dynamical structure factor $S(k,\omega)$
for (a) charge excitation and (b)
spin excitation for the system
$L=14$, $J/t=1$ and $12/14$ in Fig. 3 (b).

Fig. 5: Velocity of the
charge-excitation spectrum for the system $L=14$, $n=12/14$
and
$J/t$ varied.

Fig. 6: Inverse of the exponent $K_\rho$ of the charge sector
for the system $L=14$, $n=12/14$ and $J/t$ varied.

Fig. 7: Phase diagram of the charge sector.
The contours of the exponent $K_\rho$ are
also shown. The plots $+$, $\times$
and $\diamond$ denote the data estimated for  the
systems of the length $L=10$, $12$ and $14$, respectively.

Fig. 8: Qualitatively different phase diagrams of the charge sector reported
so far. The type (a) was reported for the $S=1/2$ $t$-$J$ model with gapless
spin excitation \cite{Ogata91a},
while the type (b) was reported for the frustrated $t$-$J$-$J'$
model with spin gap \cite{Ogata91b}.

Fig. 9: String correlation ${\cal O}^z_{\rm string}(L/2)$ for the system
with $J/t=1$, and $n$ varied.

Fig. 10 (a) (b) (c): Binder parameter $U(L)$ of the string correlation
for the systems with
(a) $J/t=0.2$, (b) $J/t=1.2$ and (c) $J/t=2.2$, and $n$ varied.

Fig. 11: Schematic drawing of the phase diagram of the spin sector.

Fig. 12 (a) (b): Generalized string correlation
${\cal O}^z_{\rm string}(L/2,\theta)$ against the angle $\theta$.
It is calculated for the systems with (a) $L=14$, $J/t=0.5$ and $n=12/14$,
and (b) $L=14$, $J/t=1.2$ and $n=4/14$.

\end{normalsize}

\begin{thebibliography}{99}
\bibitem{Rossat91a}J. Rossat-Mignod, L. P. Regnault, C. Vettier, P. Burlet,
J. Y. Henry and G. Lapertot: Physica {\bf B 169} (1991) 58.
\bibitem{Rossat91b}J. Rossat-Mignod, L. P. Regnault, C. Vettier, P. Burlet,
P. Burlet, J. Bossy, J. Y. Henry and G. Lapertot:
Physica {\bf C 185-189} (1991)
86.
\bibitem{Imada93}M. Imada: Phys. Rev. {\bf B 48} (1993) 550.
\bibitem{Ogata91b}M. Ogata, M. U. Luchini and T. M. Rice: Phys. Rev. {\bf B 44}
12083.
\bibitem{Dagotto92}E. Dagotto, J. Riera and D. J. Scalapino:
Phys.\ Rev.\ {\bf B 45} (1992) 5744.
\bibitem{Johnston87}D. C. Johnston, J. W. Johnson,
D. P. Goshorn and A. J. Jacobson: Phys.\ Rev.\ B{\bf 35} (1987)
219.
\bibitem{Takano92}M. Takano, Z. Hiroi M. Azuma and Y.
Takeda: Jpn.\ J. Appl.\ Phys.\ Ser.\ {\bf 7}
(1992) 3.
\bibitem{Scalapino95}D. J. Scalapino: Nature Vol. 377 (1995) 12.
\bibitem{Hiroi95}Z. Hiroi and M. Takano: Nature Vol. 377 (1995) 41.
\bibitem{Dagotto95}E. Dagotto and T. M. Rice: to be published in Science.
\bibitem{Haldane83a}F. D. M. Haldane: Phys.\ Lett.\ {\bf 93A} (1983) 464.
\bibitem{Haldane83b}F. D. M. Haldane: Phys.\ Rev.\ Lett.\ {\bf 50} (1983) 1153.
\bibitem{Kolb83}M. Kolb, R. Botet and R. Jullien: J. Phys. A: Math. Gen.
{\bf 16} (1983) L673.
\bibitem{Buyers86}W. J. L. Buyers, R. M. Morra, R. L. Armstrong, M. J.
Hogan, P. Gerlach and K. Hirakawa: Phys. Rev. Lett. {\bf 56} (1986) 371.
\bibitem{Batlogg94}B. Batlogg, S-W. Cheong, L. W. Rupp, Jr: Physica
{\bf B 194-196 } (1994) 173.
\bibitem{Ramirez94}A. P. Ramirez, S-.-W. Cheong and M. L. Kaplan:
Phys. Rev. Lett. {\bf 72} (1994) 3108.
\bibitem{Di94}J. F. DiTusa, S-W. Cheong, J.-H. Park, G. Aeppli, C.
Broholm and C. T. Chen: Phys. Rev. Lett. {\bf 73} (1994) 1857.
\bibitem{Penc95}K. Penc and H. Shiba: Phys. Rev. {\bf B 52} (1995) R715.
\bibitem{Dagotto95b}E. Dagotto, J. Riera, A. Sandvik and A. Moreo: preprint.
\bibitem{Koshibae95}W. Koshibae, Y. Ohta and S. Maekawa: preprint.
\bibitem{Kaburagi94}M. Kaburagi and T. Tonegawa: Jour. Phys. Soc Jpn. {\bf 63}
(1994) 420.
\bibitem{Sorensen95}E. S. S{\o}rensen and I. A. Affleck: Phys. Rev. {\bf B 51}
(1995) 16115.
\bibitem{Lu95}Z-Y. Lu, Z-B. Su and L. Yu: Phys. Rev. Lett. {\bf 74} (1995)4297.
\bibitem{Tonegawa95}T. Tonegawa and M. Kaburagi: Jour. Phys. Soc. Jpn.
{\bf 64} (1995) 3956.
\bibitem{White93}S. R. White and D. A. Huse: Phys. Rev. {\bf B 48} (1993) 3844.
\bibitem{Golinelli94}O. Golinelli, Th. Jolic{\oe}ur and R. Lacaze:
Phys. Rev. {\bf B 50} (1994) 3037.
\bibitem{Zhang89}S. Zhang and D. P. Arovas: Phys. Rev. {\bf B 40} 2708.
\bibitem{Ogata91a}M. Ogata, M. U. Luchini, S. Sorella and F. F. Assaad:
Phys. Rev. Lett. {\bf 66} (1991) 2388.
\bibitem{AKLT87}I. Affleck, T. Kennedy, E. H. Lieb and H. Tasaki: Phys.\ Rev.\
Lett.\ {\bf 59} (1987) 799.
\bibitem{AKLT88}I. Affleck, T. Kennedy, E. H. Lieb and H. Tasaki: Commun.\
Math.\ Phys.\ {\bf 115} (1988) 477.
\bibitem{Arovas88}D. P. Arovas, A. Auerbach and F. D. M. Haldane:
Phys. Rev. Lett. {\bf 60} (1988) 531.
\bibitem{den89}M. den Nijs and K. Rommelse: Phys.\ Rev.\ B{\bf 40} (1989) 4709.
\bibitem{Tasaki91}H. Tasaki: Phys.\ Rev.\ Lett.\ {\bf 66} (1991) 98.
\bibitem{Schulz91}H. J. Schulz: {\it
Proceedings of Adriatico Research Conference
and Workshop (July 1990)-Strongly Correlated Electron Systems II}
(World Scientific, Singapore, 1991), p. 57
\bibitem{Mori65}H. Mori: Prog. Theor. Phys {\bf 33} (1965) 423.
\bibitem{Gagliano87}E. R. Gagliano and C. A. Balseiro: Phys. Rev. Lett.
{\bf 59} (1987) 2999.
\bibitem{Bares90}P. A. Bares and G. Blatter: Phys. Rev. Lett. {\bf 64}
(1990) 2567.
\bibitem{Haldane81}F. D. M. Haldane: J. Phys. {\bf C 14} (1981) 2585.
\bibitem{Schulz90}H. J. Schulz: Phys. Rev. Lett. {\bf 64} (1990) 2831.
\bibitem{Frahm90}H. Frahm and V. E. Korepin: Phys. Rev. {\bf B 42}
(1990) 10553.
\bibitem{Kawakami90}N. Kawakami and S. K. Yang: Phys. Rev. Lett {\bf 65}
(1990) 2309.
\bibitem{Majumdar69}C. K. Majumdar and D. K. Ghosh: J. Math. Phys. {\bf 10}
(1969) 1399.
\bibitem{Binder81a}K. Binder: Phys.\ Rev.\ Lett.\ {\bf 47} (1981) 693.
\bibitem{Binder81b}K. Binder:  Z. Phys.\ {\bf B}-Cond.\ Matt.\ {\bf 43} 119.
\bibitem{Hatano94}N. Hatano: J. Phys.\ A: Math.\ Gen.\ {\bf 27} (1994) L223.
\bibitem{Oshikawa92}M. Oshikawa: J. Phys.: Condens, Matter {\bf 4} 7469.
\bibitem{Totsuka95}K. Totsuka and M. Suzuki: J. Phys.: Condens. Matter
{\bf 7} (1995) 1639.
\bibitem{Nishiyama95}Y. Nishiyama, K. Totsuka, N. Hatano and M. Suzuki:
Jour. Phys. Soc. Jpn. {\bf 64} (1995) 414.
\bibitem{Totsuka95b}K. Totsuka, Y. Nishiyama, N. Hatano and M. Suzuki:
J. Phys.: Condens. Matter {\bf 7} (1995) 4895.
\bibitem{Hellberg93}C. S. Hellberg and E. J. Mele: Phys. Rev. {\bf B 48}
(1993) 646.

\end{thebibliography}
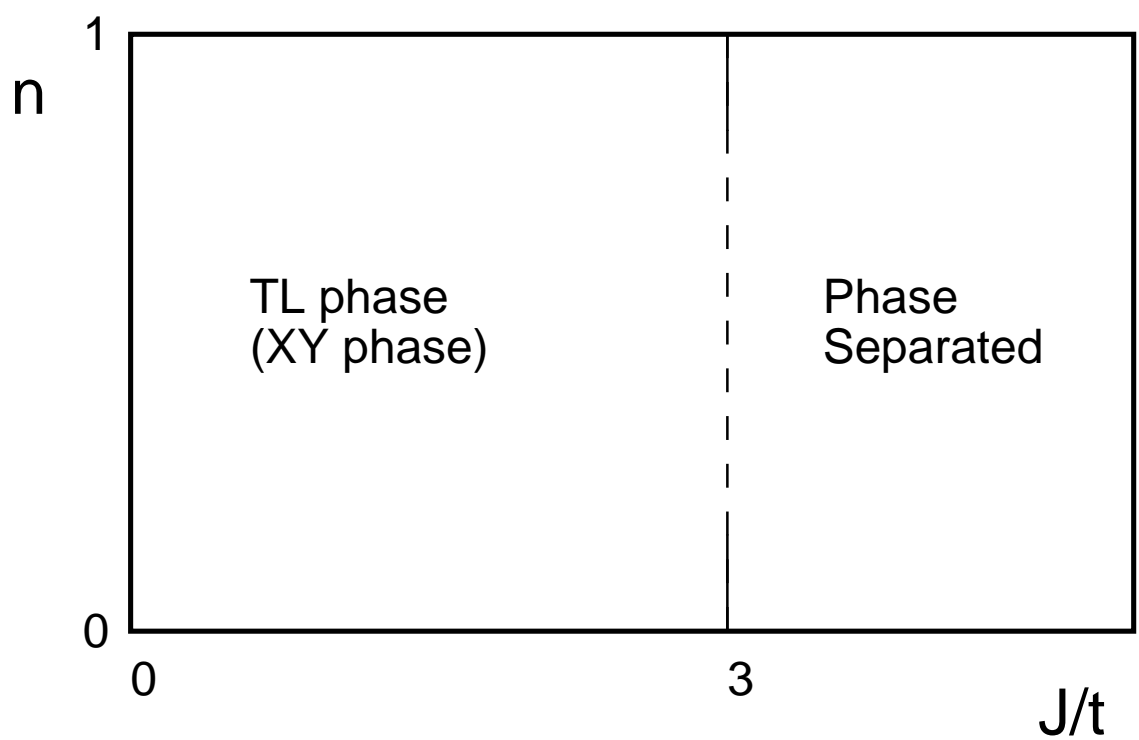
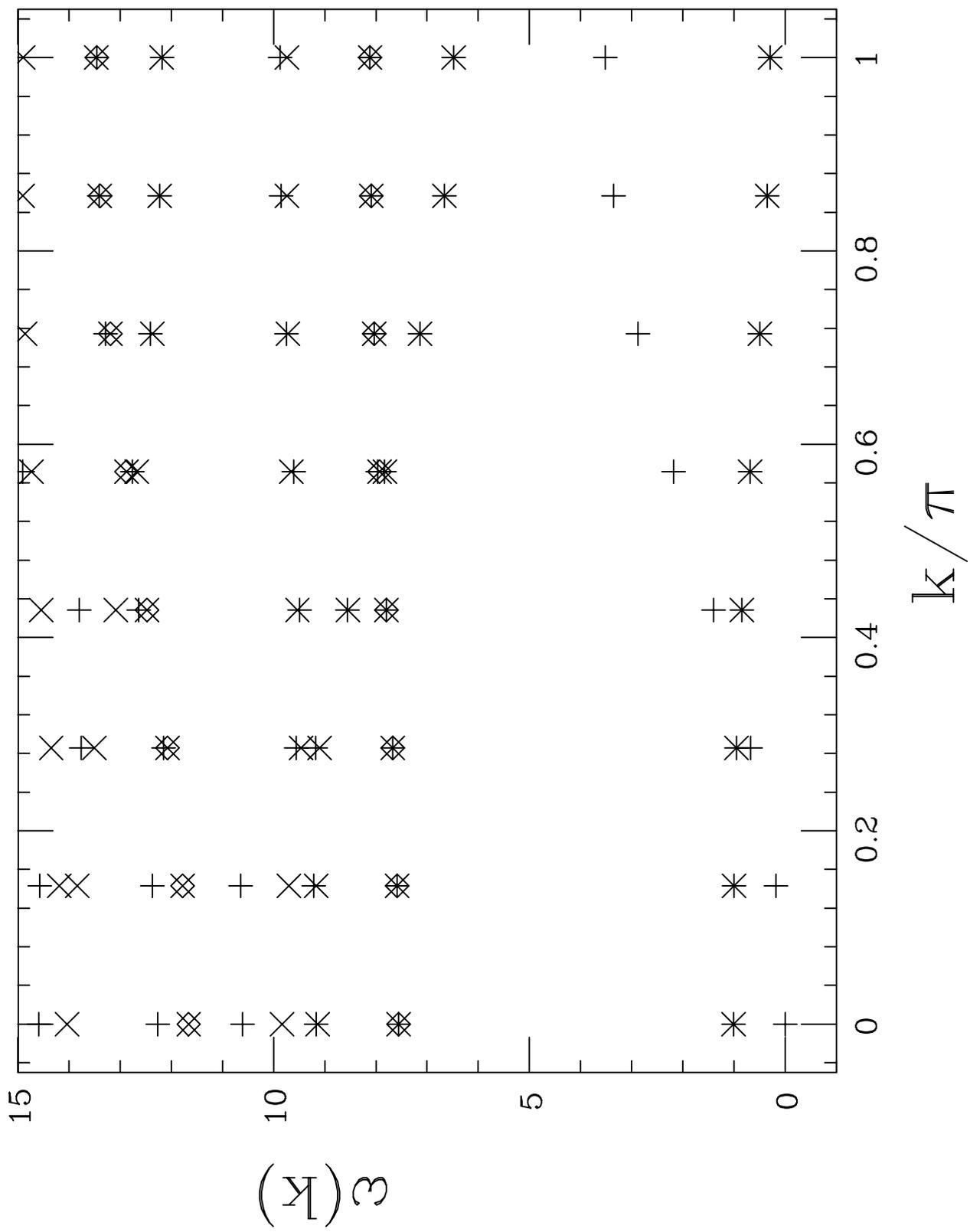
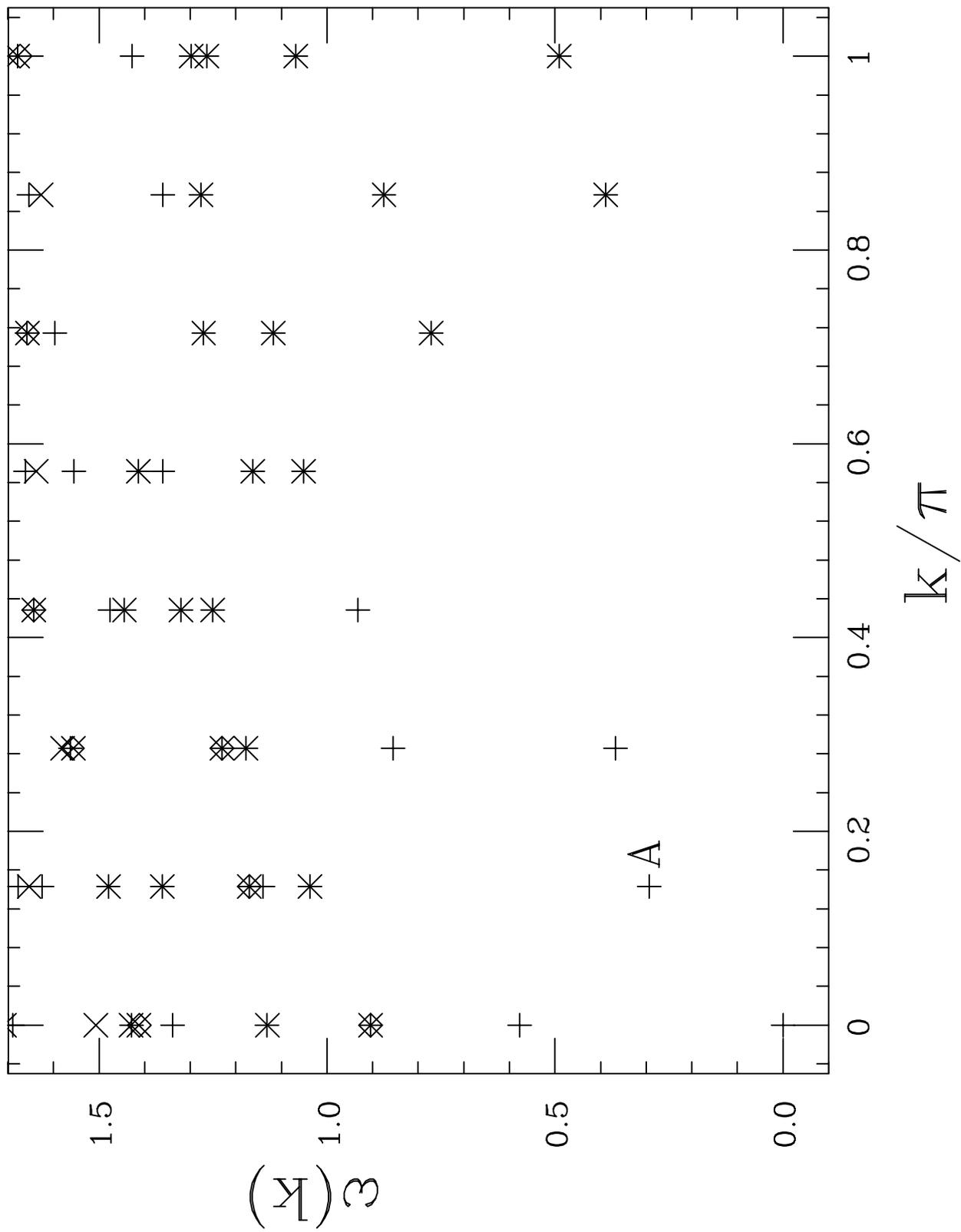
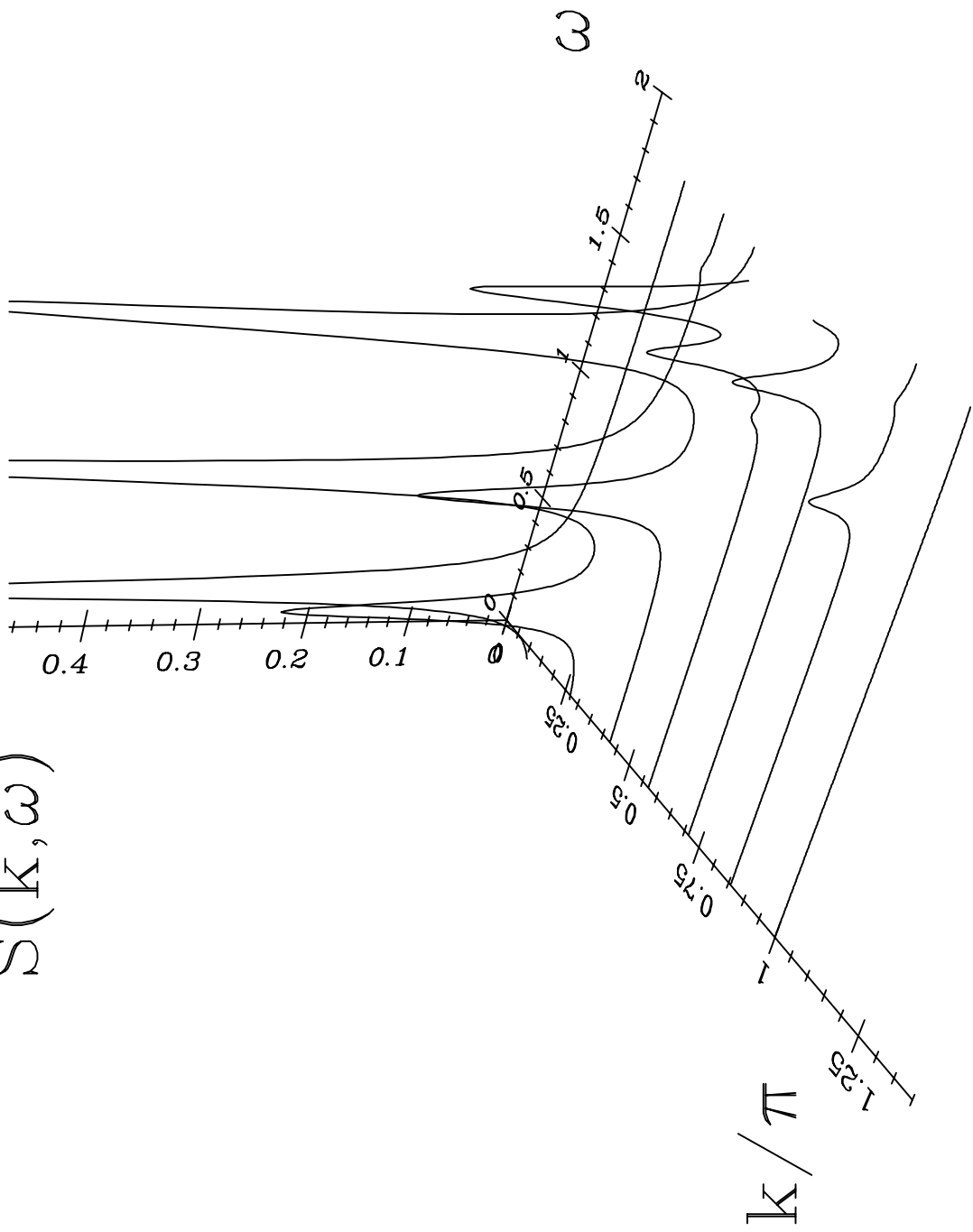
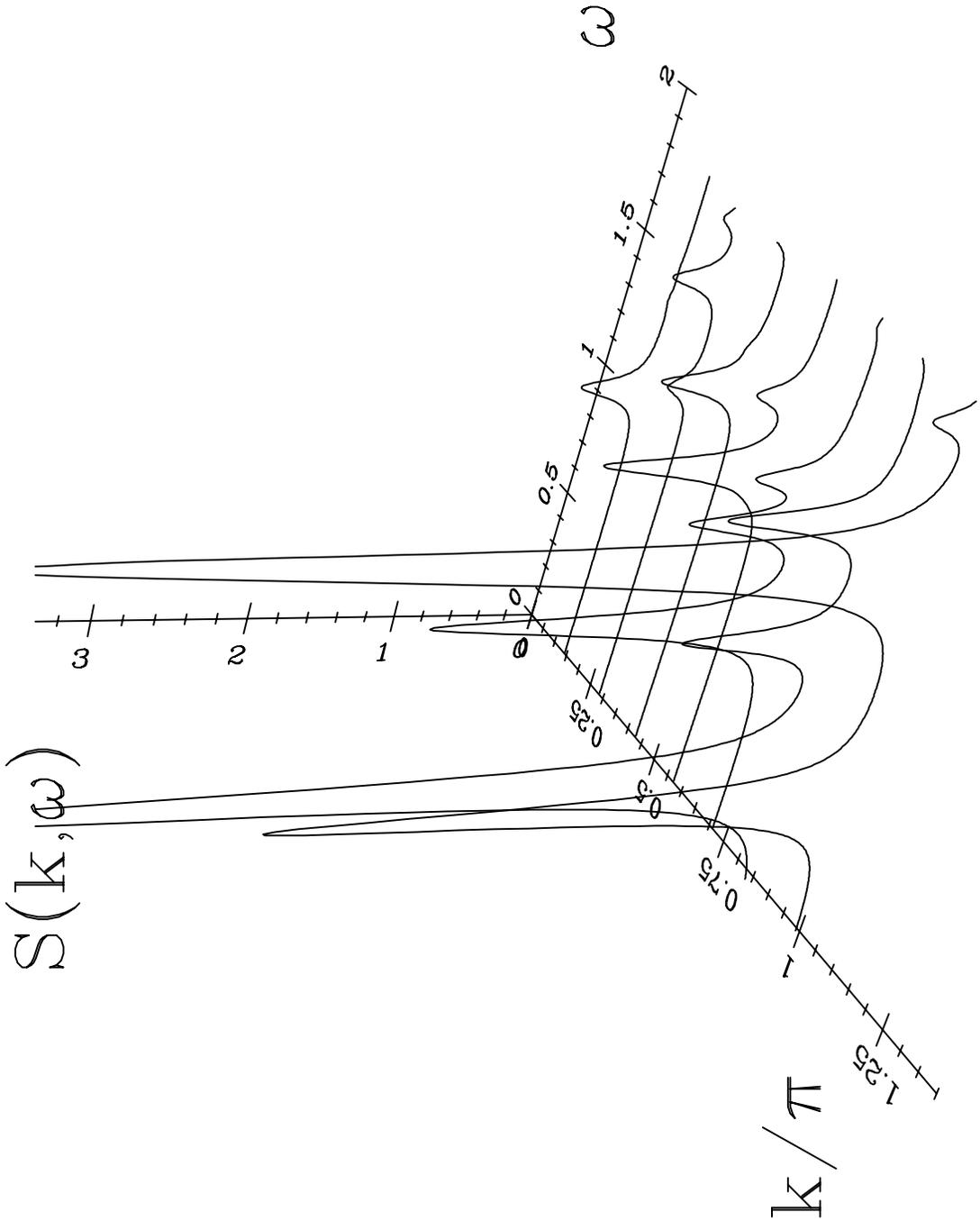
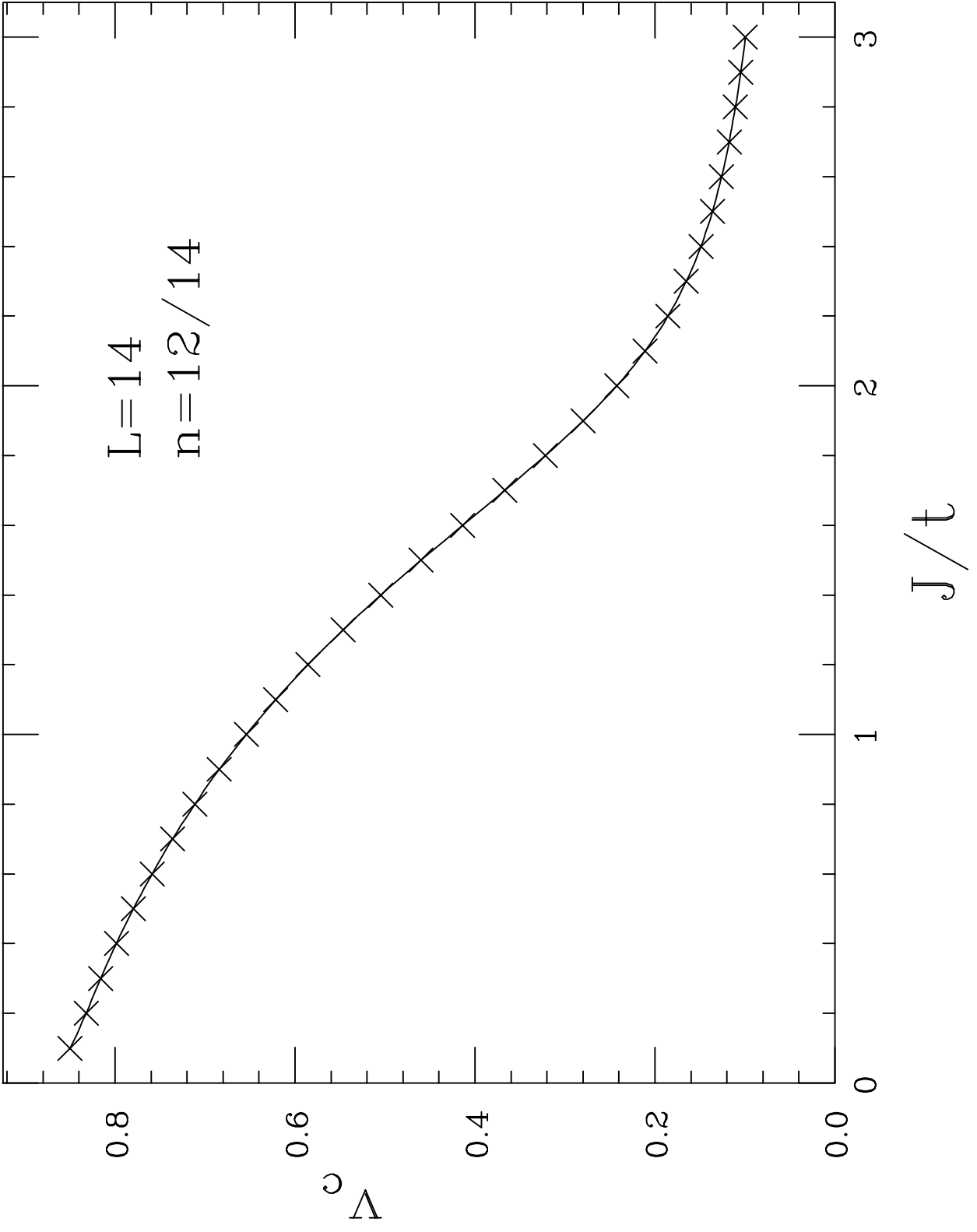
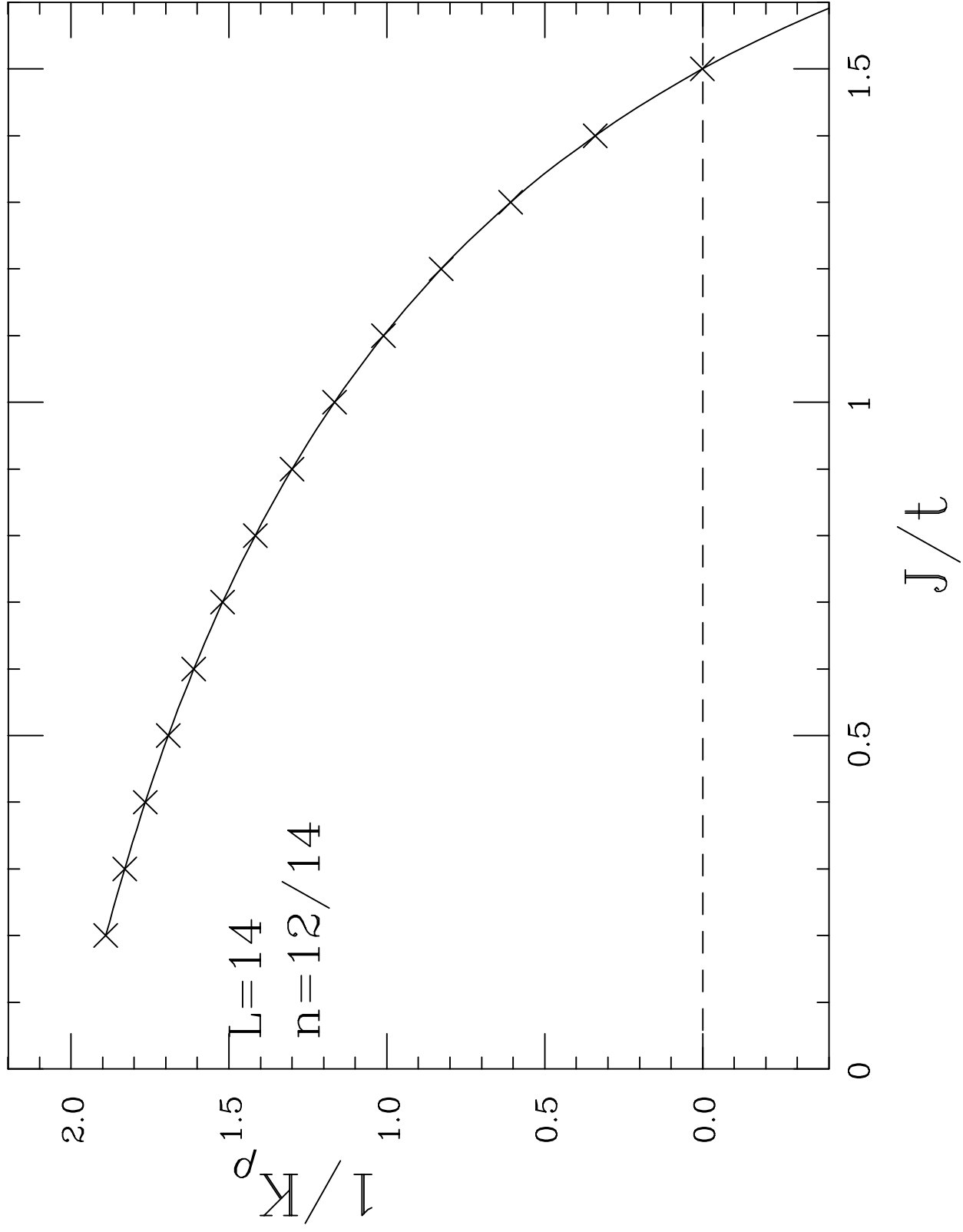
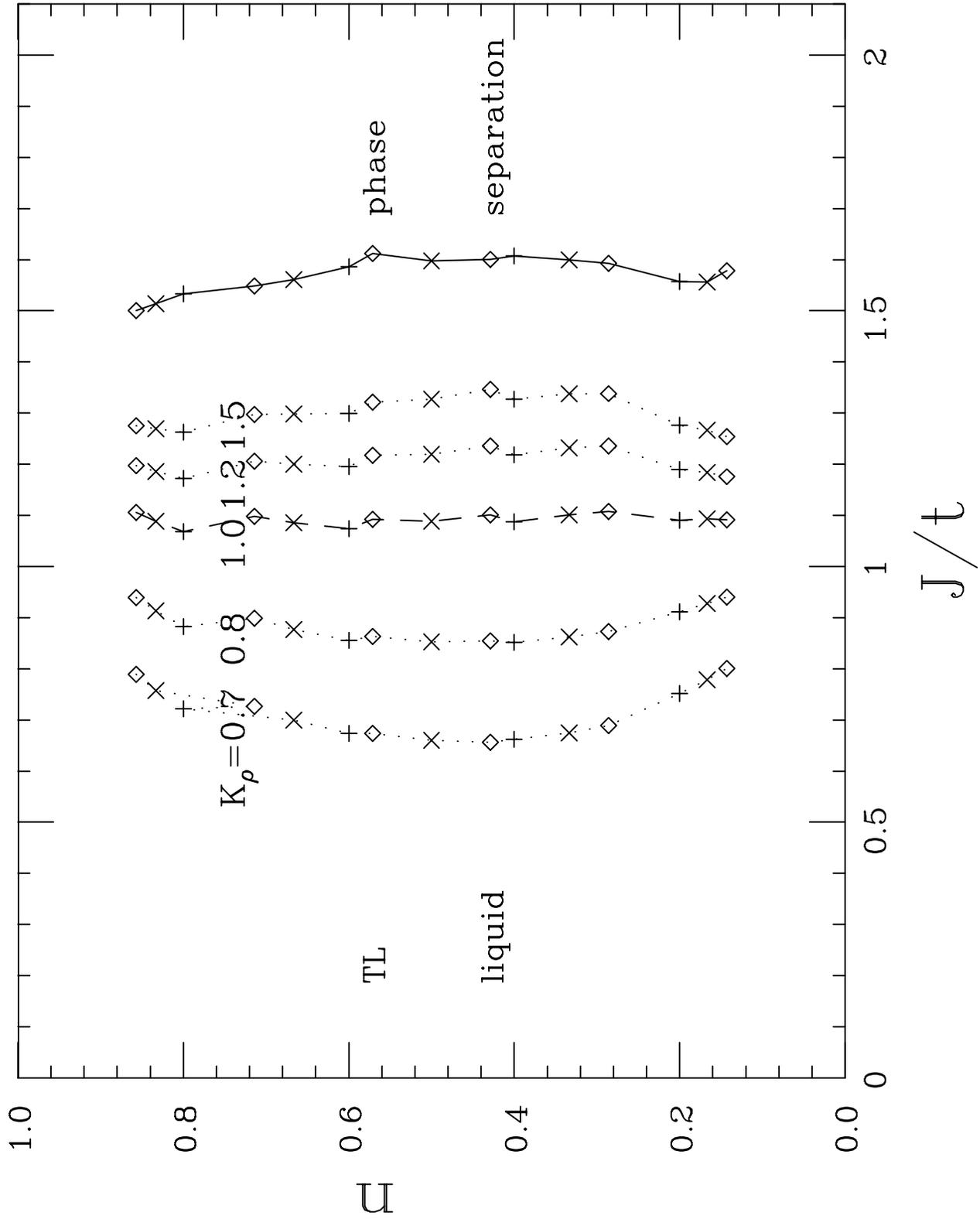
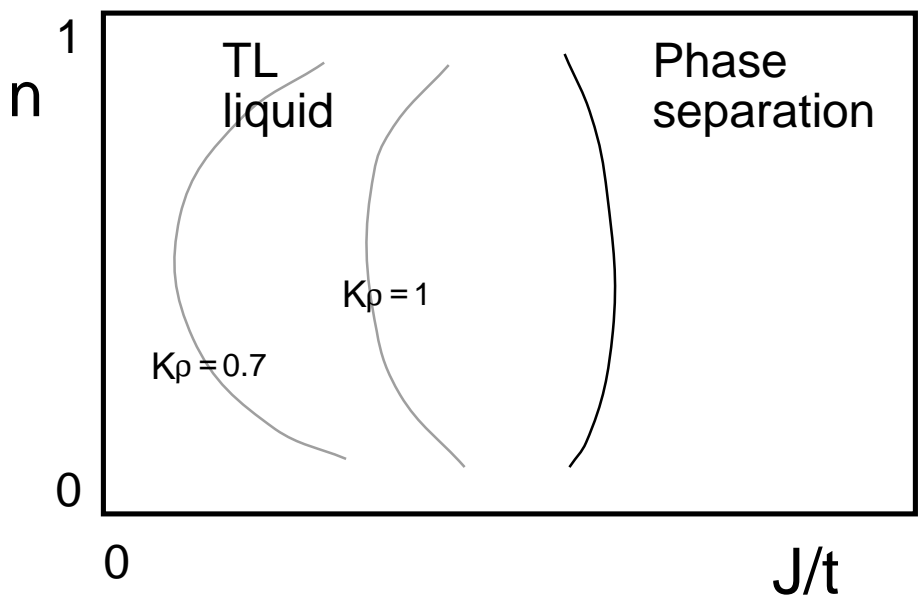
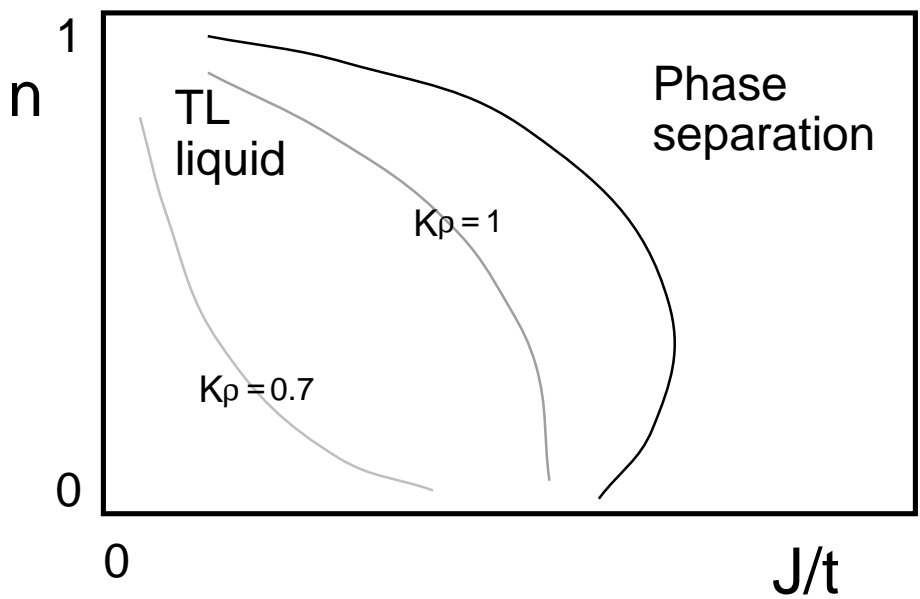
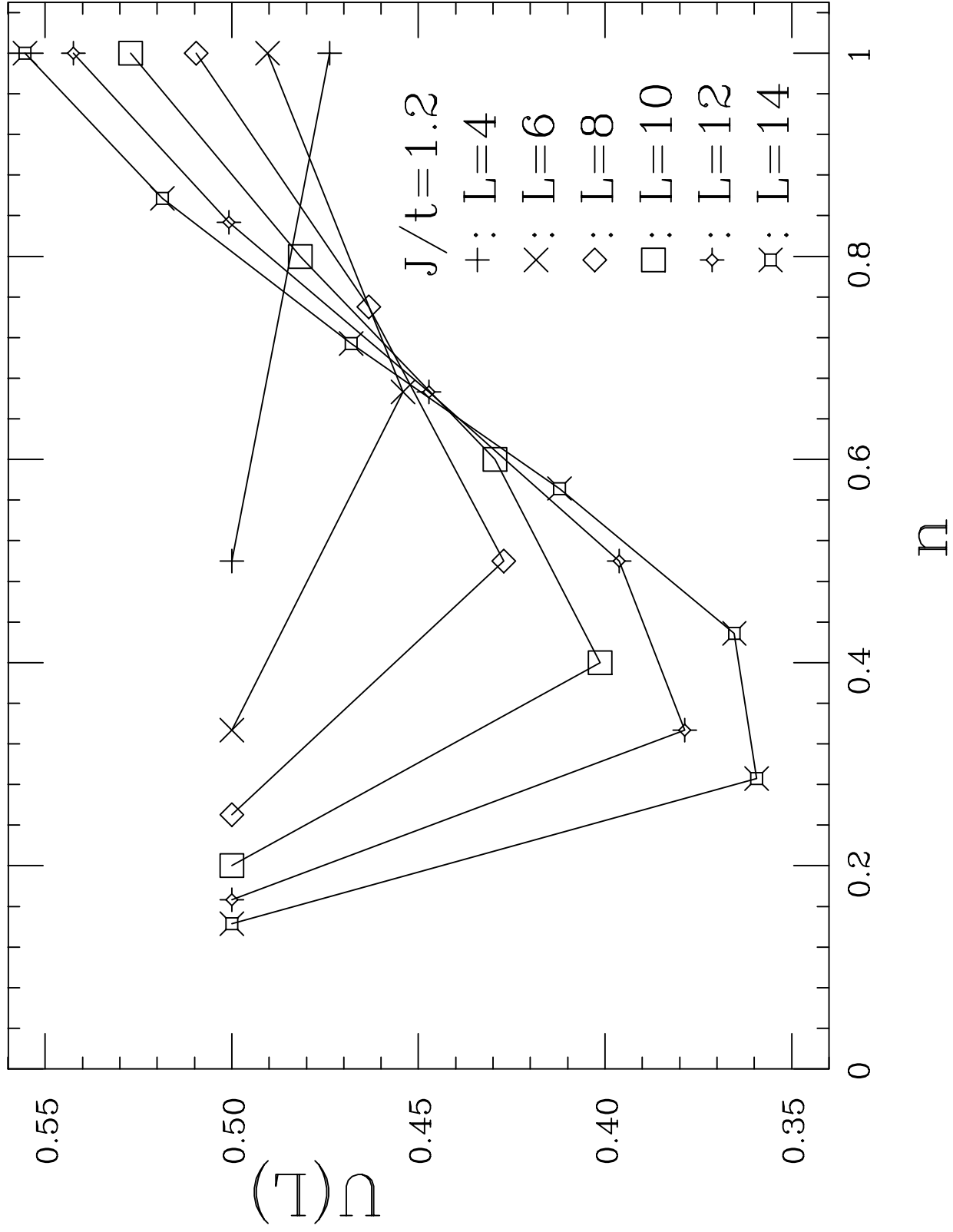
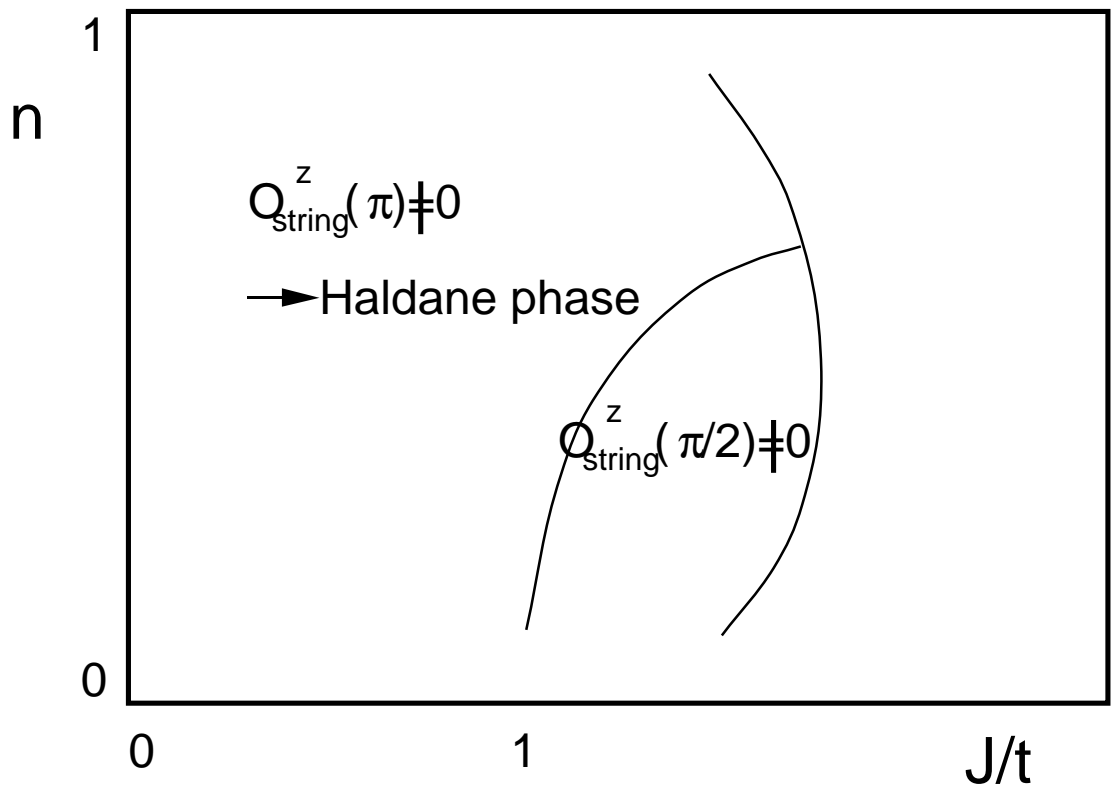
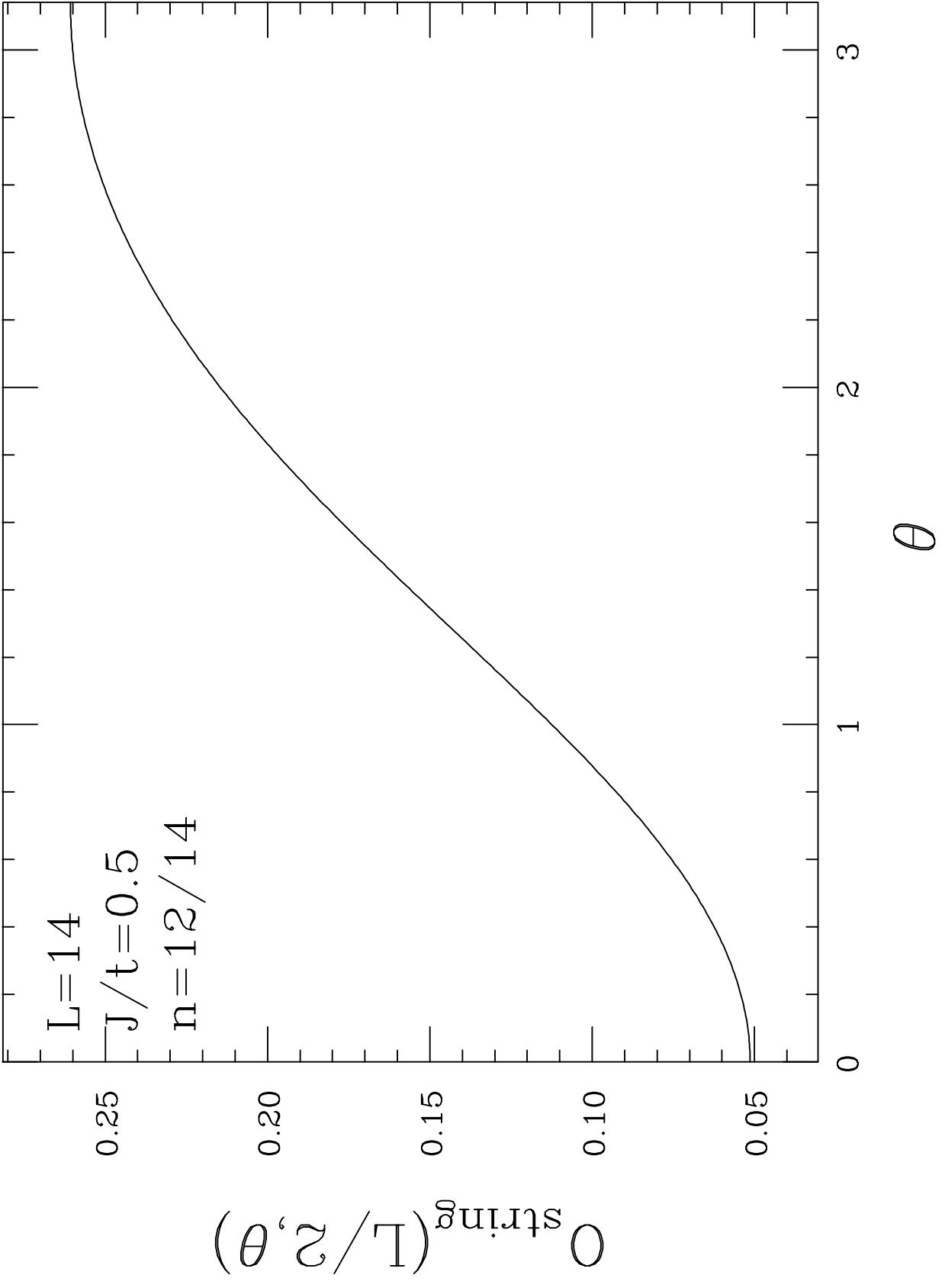
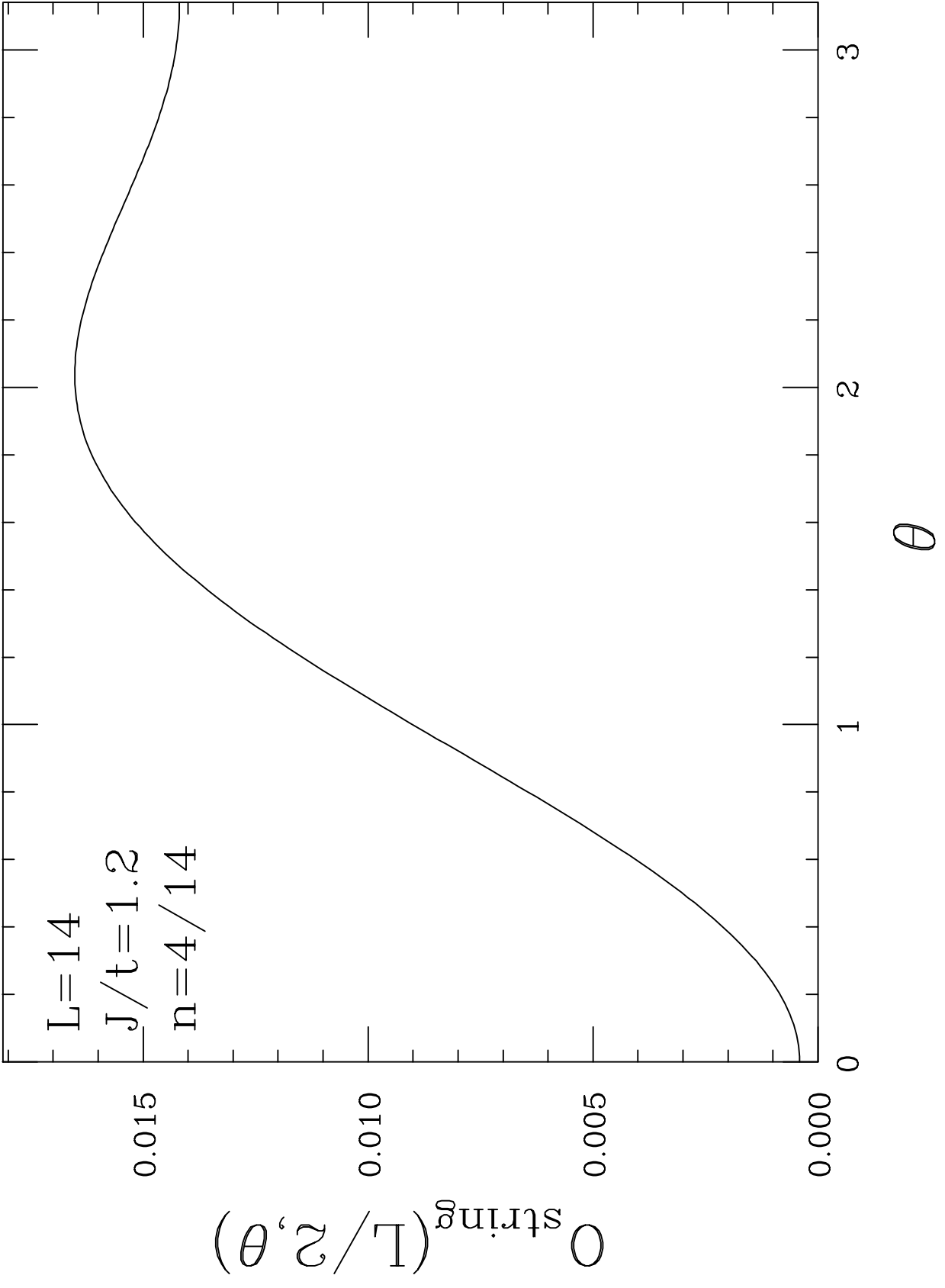
\end{document}